\documentclass[11pt]{article}
\usepackage{jheppub}
\usepackage{bbm} 
\allowdisplaybreaks 

\newcommand{\la}[1]{\label{#1}}

\newcommand{\eq}{Eq.~}
\newcommand{\se}{Sec.~}
\newcommand{\app}{App.~}
\newcommand{\eqs}{Eqs.~}
\newcommand{\nr}[1]{(\ref{#1})}
\newcommand{\nn}{\nonumber \\}
\renewcommand{\(}{\left(}
\renewcommand{\)}{\right)}
\newcommand{\lb}{\left\{}
\newcommand{\rb}{\right\}}
\newcommand{\lk}{\left[}
\newcommand{\rk}{\right]}

\newcommand{\e}{\epsilon}
\newcommand{\order}[1]{{\cal O}(#1)}

\newcommand{\sumint}[1]{\hbox{$\sum$}\!\!\!\!\!\!\!\int_{#1}}
\newcommand{\sumintp}[1]{\hbox{$\sum^\prime$}\!\!\!\!\!\!\!\!\!\int_{#1}}
\newcommand{\sint}{\sum\!\!\!\!\!\!\int\;}

\newcommand{\gammaE}{{\gamma_{\small\rm E}}}
\newcommand{\gammaEs}{{\gamma}}
\newcommand{\csch}[0]{\text{csch}}
\newcommand{\mE}{m_{\mathrm{E}}}
\newcommand{\intV}{{\cal M}_{1,0}}
\renewcommand{\vec}[1]{{\mathbf{#1}}}
\newcommand{\pz}{P_0}
\newcommand{\pzb}{\bar{P}_0}
\newcommand{\Pz}{P_0}
\newcommand{\qz}{Q_0}
\newcommand{\Qz}{Q_0}
\newcommand{\Rz}{R_0}
\newcommand{\dpzero}{\delta_{\pz}}
\newcommand{\dqzero}{\delta_{\qz}}
\newcommand{\intx}{\int_0^\infty\!\!\!\!\!{\rm d}x}
\newcommand{\intxx}{\int_0^\infty\frac{{\rm d}x}{x}}
\newcommand{\inty}{\int_0^\infty\!\!\!\!\!{\rm d}y}
\newcommand{\Vb}{V_2}
\newcommand{\cS}{{\cal V}}
\newcommand{\sS}{V}

\title{A new three-loop sum-integral of mass dimension two}

\preprint{BI-TP 2012/30}

\author{Ioan Ghi\c{s}oiu}
\author{and York Schr\"oder}

\affiliation{Faculty of Physics, University of Bielefeld,
  33501 Bielefeld, Germany}

\emailAdd{ighisoiu@physik.uni-bielefeld.de}
\emailAdd{yorks@physik.uni-bielefeld.de}

\abstract{We evaluate a new 3-loop sum-integral
which contributes to the Debye screening mass 
in hot QCD. While we manage to derive all divergences
analytically, its finite part is mapped onto 
simple integrals and evaluated numerically.}

\begin{document}
\maketitle
\flushbottom

%
\section{Introduction}
\la{se:intro}

Motivated by the need to compute a number of unknown 
three-loop vacuum sum-integrals of mass dimension $2$,
in order to complete the evaluation of the 3-loop Debye screening
mass $\mE^2$ in hot QCD \cite{Moeller:2012da}, 
we turn our attention towards one particular open case.
The present paper is hence rather technical in practice,
but serves to establish one of the few missing building
blocks for the determination of the Debye mass at NNLO.

In Ref.~\cite{Moeller:2012da}, 
the bosonic contribution to $\mE^2$ has been reduced
to a sum of ten sum-integrals times pre-factors which are rational 
functions in $d$,
by systematic \cite{Laporta:2001dd} use 
of integration-by-parts (IBP) relations \cite{Chetyrkin:1981qh},
generalized to finite temperatures \cite{Nishimura:2012ee}.
Two out of the basis of those ten master sum-integrals were
triple products of 1-loop cases $I$ and hence 
trivial (cf.\ \eq\nr{eq:Idef}), while only two further 
non-trivial three-loop
representatives of that basis (containing basketball-type sum-integrals) 
are presently known, 
namely $S_1$ of \cite{Gynther:2007bw} 
(see also \cite{Schroder:2008ex,Moller:2010xw}) 
and $B_{3,2}$ of \cite{Moeller:2012da},
however not deeply enough in their $\e$\/-expansion 
since the corresponding pre-factors develop single poles as 
$d\!\rightarrow\!3$.

It can be shown that, by a change of basis,
the eight non-trivial three-loop basketball-type masters 
can be traded for only three spectacles-type (plus some trivial
factorized) sum-integrals, with pole-free pre-factors \cite{debyeMass}.
Of those three new masters, it turns out that one has already
appeared in a study of the $g^7$ contributions to the pressure of
massless thermal scalar $\phi^4$ theory in the framework of screened
perturbation theory. It has originally been computed 
in Ref.~\cite{Andersen:2008bz}
and been re-derived in Ref.~\cite{Schroder:2012hm}, 
where it was named $\intV$.
This leaves us with two elements of the new basis to be evaluated,
one of which we tackle in the present paper.

Let us define
the massless bosonic 3-loop vacuum sum-integral $\Vb$ 
in terms of the 1-loop 2-point sum-integrals $\Pi,\bar\Pi$ as 
\begin{align}
\Vb &\equiv \sumint{P} \frac1{\lk P^2\rk^2}\,\Pi(P)\,\bar\Pi(P)
\;,\quad 
\Pi(P) = \sumint{Q} \frac1{Q^2\,(P-Q)^2}
\;,\quad 
\bar\Pi(P) = \sumint{R} \frac{\Rz^2}{R^2\,(P-R)^2}
\;.
\end{align}
We use (Euclidean) bosonic four-momenta 
$P=(\pz,\vec{p})=(2\pi n_p T,\vec{p})$
with $P^2 = \pz^2 + \vec{p}^{2}$;
the temperature of the system is denoted by $T$;
and the sum-integral symbol
is a shorthand for
\begin{align}
\sumint{P} \equiv
T\sum_{n_p\in\mathbbm{Z}}\int\frac{\mathrm{d}^{d}\vec p}{(2\pi)^{d}}\;,
\quad\mbox{with~~}d=3-2\e \;.
\end{align}
In the remainder of this paper, we evaluate the new sum-integral
$\Vb$ analytically as far as possible, utilizing methods from
\cite{Arnold:1994ps,Gynther:2007bw,Moller:2010xw,Schroder:2012hm}. 
In \se\ref{se:decomposeVb} we split up the sum-integral into nine
pieces that are either calculable analytically, or explicitly
finite such that they can be evaluated numerically in $d\!=\!3$.
\se\ref{se:evalVb} treats each of these nine terms in turn,
which are then summed to obtain our final result in \se\ref{se:resultVb}.
We conclude in \se\ref{se:conclu},
while some technicalities are relegated to the appendices.

%
\section{Decomposition of $\Vb$}
\la{se:decomposeVb}

Inspired by \cite{Arnold:1994ps}
and guided by our experience from \cite{Schroder:2012hm},
$\Vb$ can be identically re-written as
\begin{align}
\la{eq:V2rewritten}
\Vb &=\,\sumint{P} \frac{\dpzero}{\lk P^2\rk^2}\,\Pi(P)\,\bar\Pi(P)
+\sumintp{P} \frac1{\lk P^2\rk^2} \Big\{ 
\lk\Pi-\Pi_B\rk\lk\bar\Pi-\bar\Pi_B\rk
+\nn&
+\Pi_D\lk\bar\Pi-\bar\Pi_C\rk
+\lk\Pi_B-\Pi_D\rk\lk\bar\Pi-\bar\Pi_C\rk
+\lk\Pi-\Pi_C\rk\bar\Pi_D
+\lk\Pi-\Pi_C\rk\lk\bar\Pi_B-\bar\Pi_D\rk
+\nn&
+\lk\Pi_C-\Pi_B\rk\bar\Pi_B
+\Pi_B\lk\bar\Pi_C-\bar\Pi_B\rk
+\Pi_B\bar\Pi_B
\Big\} 
\;,\\
\la{eq:V1to9}
&\equiv \cS_1+\cS_2+\cS_3+\cS_4+\cS_5+\cS_6+\cS_7+\cS_8+\cS_9\;,
\end{align}
where $\dpzero$ picks out the Matsubara zero-mode, 
the primed sum excludes the zero-mode
and we have suppressed the argument $(P)$ of all functions 
in curly brackets.
Our strategy amounts to choosing
($\Pi_B$: zero-T part of $\Pi$; 
$[\Pi_C-\Pi_B]$: large-P behavior of remainder)
\begin{align}
\la{eq:PiBCD}
\Pi_B = \int_Q\frac1{Q^2(P-Q)^2} = \frac{G(1,1,d+1)}{(P^2)^{(3-d)/2}}\;,\quad
\Pi_C = \Pi_B+\frac{2\,I_1^0}{P^2}\;,\quad
\Pi_D = \frac{G(1,1,d+1)}{(\alpha_1\,T^2)^{(3-d)/2}}\;,
\end{align}
where the functions $G$ and $I$ are known analytically as
given in \app\ref{se:functions} while  
$\alpha_1$ (and $\alpha_2$ below) are constants to be 
fixed later, as well as
(with $U=(1,\vec 0)$)
\begin{align}
\bar\Pi_B &= \int_R\frac{\Rz^2}{R^2(P-R)^2} 
= U_\mu U_\nu \int_R\frac{R_\mu R_\nu}{R^2(P-R)^2}
= U_\mu U_\nu \lb g_{\mu\nu}A(P^2)+P_\mu P_\nu B(P^2)\rb\nn
&= A(P^2)+\Pz^2 B(P^2)
= \frac{(d+1)\Pz^2-P^2}{4d}\,\Pi_B\;,\\
\bar\Pi_C &= \bar\Pi_B+\frac{\Pz^2 I_1^0+2I_1^2}{P^2}\;,\quad
\bar\Pi_D = \frac{(d+1)\Pz^2-P^2}{4d}\,
\frac{G(1,1,d+1)}{(\alpha_2\,T^2)^{(3-d)/2}}
\;.
\end{align}

%
\section{Evaluation of the nine terms of \eq\nr{eq:V2rewritten}}
\la{se:evalVb}

We will now turn to each term of \eq\nr{eq:V2rewritten} in sequence,
evaluating it up to the constant part.

%
\subsection{Evaluation of $\cS_1$}
\la{se:1st}

To tame the infrared behavior of 
the first term in \eq\nr{eq:V2rewritten}, 
we transform it via the integration-by-parts (IBP)
relation of \eq\nr{eq:ibp1} into
\begin{align}
\cS_1&\equiv
\sumint{P} \frac{\dpzero}{\lk P^2\rk^2}\,\Pi(P)\,\bar\Pi(P)
= \frac1{d-6} \sumint{P} \frac{\dpzero}{P^2} \lb
\widetilde\Pi\,\bar\Pi
+\Pi\,\bar{\bar\Pi}
-I_2^0\frac1{P^2}\,\bar\Pi
-I_2^2\frac1{P^2}\,\Pi
\rb
\end{align}
with
\begin{align}
\widetilde\Pi= \sumint{Q} \frac1{[Q^2]^2\,(P-Q)^2}\;,\quad
\bar{\bar\Pi}= \sumint{R} \frac{\Rz^2}{[R^2]^2\,(P-R)^2}\;.
\end{align}
Using $\Pi_A=\Pi_B+\frac{T\,G(1,1,d)}{(P^2)^{2-d/2}}$ 
and $\bar{\bar\Pi}_B=\int_R \frac{\Rz^2}{[R^2]^2\,(P-R)^2}$
as well as the functions $G$, $I$ and $A$ that are listed 
in \app\ref{se:functions}, this is then identically re-written as
\begin{align}\la{eq:S9decomp}
\cS_1 =&\,
\frac1{d-6} \sumint{P} \frac{\dpzero}{P^2} \Big\{
\bar\Pi_B\widetilde\Pi
+\lk\bar\Pi-\bar\Pi_B\rk\sumint{Q} \frac{\dqzero}{[Q^2]^2\,(P-Q)^2}
+\lk\bar\Pi-\bar\Pi_B\rk\sumintp{Q} \frac1{[Q^2]^2\,(P-Q)^2}
+\nn&
+\lk\Pi-\Pi_A\rk\lk\bar{\bar\Pi}-\bar{\bar\Pi}_B\rk
+\Pi\,\bar{\bar\Pi}_B
+\Pi_A\bar{\bar\Pi}
-\Pi_A\bar{\bar\Pi}_B
-I_2^0\frac1{P^2}\,\bar\Pi
-I_2^2\frac1{P^2}\,\Pi
\Big\}\\
=&\,\frac1{d\!-\!6} \Big\{ 
-\frac{G(1,1,d\!+\!1)}{4d}\,A(\tfrac{3-d}2,2,1;0)
+\Big[T\,G(2,1,d)\,A(\tfrac{8-d}2,1,1;2) -0_{\rm scalefree}\Big]+\cS_{1a}
+\nn&
+\cS_{1b}+\frac{2G(1,1,d\!+\!1)-G(2,1,d\!+\!1)}{4d}\,A(\tfrac{5-d}2,1,1;0)
+\Big[G(1,1,d+1)\,A(\tfrac{5-d}2,2,1;2)
+\nn&
+T\,G(1,1,d)\,A(\tfrac{6-d}2,2,1;2)\Big]
-0_{\rm scalefree}
-I_2^0\,A(2,1,1;2)
-I_2^2\,A(2,1,1;0)
\Big\}\\
\approx&\,\frac{T^2}{(4\pi)^4}\(\frac{\pi}{4 e^\gammaE}\)^\e\frac1{T^{6\e}}
\Big\{\frac1{48\e^2}+\frac{10}{48\e}+{\rm const}+\order{\e}\Big\}\;,
\end{align}
where the const also includes the numerical contributions from 
$\cS_{1a}$ and $\cS_{1b}$, 
which are defined in \eqs\nr{eq:Vb1a}, \nr{eq:Vb1b} below.
For the first term of \eq\nr{eq:S9decomp}, we have used that 
$\dpzero\bar\Pi_B=-\frac{G(1,1,d+1)}{4d}
\frac{\dpzero}{(P^2)^{(1-d)/2}}$ is a massless bubble,
while for the second term we have used that 
the $Q$\/-integration gives a massless bubble,
\begin{align*}
\dpzero\sumint{Q}\frac{\dqzero}{[Q^2]^2(P-Q)^2}&=
T\int\frac{{\rm d}^d\vec{q}}{(2\pi)^d}\,
\frac1{[\vec q^2]^2(\vec p-\vec q)^2}
= \dpzero T \frac{G(2,1,d)}{(P^2)^{3-d/2}} \;,
\end{align*}
leaving exactly solvable 2-loop tadpoles $A$.
The next two (finite) sum-integrals $\cS_{1a}$ and $\cS_{1b}$ 
will be treated numerically in coordinate space, as discussed 
below.
For the fifth and seventh terms in \eq\nr{eq:S9decomp}, we have used that 
(with $U=(1,\vec 0)$)
\begin{align*}
\dpzero\bar{\bar\Pi}_B&=
\dpzero U_\mu U_\nu \int_R\frac{R_\mu R_\nu}{[R^2]^2(P-R)^2}
=\dpzero U_\mu U_\nu \lb g_{\mu\nu}A(P^2)+P_\mu P_\nu B(P^2)\rb
=\dpzero A(P^2)
\\&
=\dpzero\frac{P^2 g_{\mu\nu}-P_\mu P_\nu}{d\,P^2}
\int_R\frac{R_\mu R_\nu}{[R^2]^2(P-R)^2}
=\frac{\dpzero}{d\,P^2}
\int_R\frac{P^2R^2-(PR)^2}{[R^2]^2(P-R)^2}
\\&
=\frac{\dpzero}{4d\,P^2} \int_R
\frac{P^2(2R^2-P^2)+(P-R)^2(P+R)^2-[R^2]^2}
{[R^2]^2(P-R)^2}
\\&
=\frac{\dpzero}{4d}\int_R\frac{2R^2-P^2}{[R^2]^2(R+P)^2}
+0_{\rm scalefree}
=\frac{\dpzero}{4d}\,
\frac{2G(1,1,d+1)-G(2,1,d+1)}{(P^2)^{\frac{3-d}2}}\;,
\end{align*}
while the remaining three terms of \eq\nr{eq:S9decomp}
are immediately seen to be 2-loop tadpoles $A$.

To complete the evaluation of $\cS_1$, we need
the two (finite) sum-integrals $\cS_{1a}$ and $\cS_{1b}$,
\begin{alignat}{2}
\la{eq:Vb1a}
\cS_{1a}
&\equiv\sumint{P} \frac{\dpzero}{P^2}
\lk\bar\Pi-\bar\Pi_B\rk\sumintp{Q} \frac1{[Q^2]^2\,(P-Q)^2}
\;\;&\approx\;\;&\frac{T^2}{(4\pi)^4}\sS_{1a}+\order{\e}\;,\\
\la{eq:Vb1b}
\cS_{1b}
&\equiv\sumint{P} \frac{\dpzero}{P^2}
\lk\Pi-\Pi_A\rk\lk\bar{\bar\Pi}-\bar{\bar\Pi}_B\rk
\;\;&\approx\;\;&\frac{T^2}{(4\pi)^4}\sS_{1b}+\order{\e}\;.
\end{alignat}

For $\sS_{1a}$, using the 3d spatial Fourier transforms 
of $[\bar\Pi-\bar\Pi_B]$ and 
$\widetilde\Pi'$ as given in \app\ref{se:FT} (used at $\pz=0$); 
integrating over $\vec{p}$ via \eq\nr{eq:av4};
letting $|\vec r|=x/(2\pi T)$ and $|\vec r'|=y/(2\pi T)$;
and using \eq\nr{eq:av5} for the angular integral:
\begin{align}
\sS_{1a}&=\frac16\intx\inty\,
\Big[\bar f(x,0)-\bar f_B(x,0)\Big] \tilde f'(y,0)\,\frac{x+y-|x-y|}{xy}\nn
&=-\frac1{36}\intxx
\Big[\bar f(x,0)-\bar f_B(x,0)\Big]
\Big[3{\rm Li}_3(e^{2x})-3\zeta(3)-2\pi^2 x+6i\pi x^2+4x^3\Big]\la{eq:numV1a}\\
&\approx -0.0285014376988(1) \;,
\end{align}
where the functions $\bar f$, $\bar f_B$ and $\tilde f'$ 
are listed in \eqs\nr{eq:FTbf}, \nr{eq:FTbfB} and \nr{eq:FTtfp}, 
respectively.

For $\sS_{1b}$, re-writing  
$\dpzero\lk\Pi-\Pi_A\rk=\dpzero\lk\Pi-\Pi_B\rk-\frac18\frac{T}{p}
\times\int\frac{{\rm d}^3\vec{r}}{r^2}\,
e^{i\vec{p}\vec{r}}\frac{8p}{(4\pi)^2} +\order{\e}$,
where $\frac18=G(1,1,3)$ while the extra integral is unity and introduced 
here for notational simplicity;
using the 3d spatial Fourier transforms of $\lk\Pi-\Pi_B\rk$ 
and $\lk\bar{\bar \Pi}-\bar{\bar \Pi}_B\rk$ (at $\pz=0$);
integrating over $\vec{p}$ via \eq\nr{eq:av4};
letting $|\vec r|=x/(2\pi T)$ and $|\vec r'|=y/(2\pi T)$;
and using \eq\nr{eq:av5} for the angular integral:
\begin{align}
\sS_{1b}&=\frac14\intx\inty\,
\Big[ f(x,0)-f_B(x,0)-1\Big]
\Big[\bar{\bar f}(y,0)-\bar{\bar f}_B(y,0)\Big]\,
\frac{x+y-|x-y|}{xy}\\
&=-\frac12\intxx
\Big[ f(x,0)-f_B(x,0)-1\Big]
\Big[ x+\ln\(x\,\csch(x)\)\Big]\la{eq:numV1b}\\
&\approx +1.197038271143294592038702(1) \;.
\end{align}

%
\subsection{Evaluation of $\cS_2$}
\la{se:2nd}

This integral is finite,
so we write 
\begin{align}
\cS_2 &\equiv \sumintp{P} \frac1{\lk P^2\rk^2} 
\lk\Pi-\Pi_B\rk\lk\bar\Pi-\bar\Pi_B\rk\approx
\frac{T^2}{(4\pi)^4}\,\sS_{2a}+\order{\e}\;.
\end{align}
Using the 3d spatial Fourier transforms of 
$[\Pi-\Pi_B]$  
and $[\bar\Pi-\bar\Pi_B]$ from \app\ref{se:FT};
using \eq\nr{eq:av1}; 
and letting $|\vec r|=x/(2\pi T)$ and $|\vec r'|=y/(2\pi T)$,
\begin{align}
\sS_{2a}=&\,\frac1{3}\intx\inty
\sum_{n=1}^\infty \frac12\int_{-1}^1\!\!\!\!{\rm d}u\,
\frac{e^{-n\sqrt{x^2+y^2+2xyu}}}{e^{n(x+y)}\;n}
\Big[f(x,n)-f_B(x,n)\Big]\Big[\bar{f}(y,n)-\bar{f}_B(y,n)\Big]\nn
\la{eq:s1234}
=&\,\frac16\intxx\int_0^x\frac{{\rm d}y}y
\sum_{n=1}^\infty
\frac{e^{-2nx}}{n^3}\Big[1+nx-ny-e^{-2ny}(1+nx+ny)\Big]\times\nn
&\,\times\Big\{\Big[f(x,n)-f_B(x,n)\Big]
\Big[\bar{f}(y,n)-\bar{f}_B(y,n)\Big]
+x\!\leftrightarrow\!y
\Big\}\;,
\end{align}
where in the last step we have 
integrated over angles via \eq\nr{eq:av2}, and split the integration region
into two to avoid taking absolute values. 
Finally, summing via \eq\nr{eq:sum1} (which gives a large 
expression containing polylogarithms)
and solving the remaining double integral over $x$ and $y$
numerically with the help of Mathematica \cite{mma}, 
\begin{align}
\sS_{2a}&\approx +0.0143560494342(1)\;.
\end{align}

%
\subsection{Evaluation of $\cS_3$}
\la{se:3rd}

Noting from \eq\nr{eq:PiBCD} 
that $\Pi_D$ is $P$\/-independent, we re-write 
the third term of \eq\nr{eq:V2rewritten} as
\begin{align}
\cS_3&\equiv
\sumintp{P}\frac1{P^4}\Pi_D\lk\bar\Pi-\bar\Pi_C\rk
=\Pi_D
\sumint{P}\frac1{P^4}\lb
\bar\Pi
-\dpzero\bar\Pi
-\bar\Pi_C
+\dpzero\bar\Pi_C
\rb\;.
\end{align}
Now the first term is a regular 2-loop sum-integral, which 
reduces via the IBP relation \eq\nr{eq:ibp5}
to a product of 1-loop tadpoles; the second term is a special
2-loop tadpole $A$; the third term is immediately recognized
to be a sum of 1-loop tadpoles by noticing that the $P$\/-dependence 
of $\bar\Pi_C$ is trivial (powers of $\Pz$ and $P^2$ only);
and, for the same reason, 
the fourth term is scale-free and hence vanishes in dimensional 
regularization.
We hence get the exact expression
\begin{align}
\cS_3=&\,
-\frac{G(1,1,d+1)}{(\alpha_1 T^2)^{(3-d)/2}}\Big\{
\frac{1}{d-5}\(2I_3^2 I_1^0+I_2^2 I_2^0\)
+A(2,1,1;2)
+\nn&+
\Big[\frac{G(1,1,d+1)}{4d}\((d+1)I_{2+\e}^2-I_{1+\e}^0\)
+I_3^2 I_1^0+2I_3^0 I_1^2\Big]
-0_{\rm scalefree}
\Big\}
\\\approx&\,\frac{T^2}{(4\pi)^4}\(\frac1{\alpha_1 T^6}\)^\e 
\Big\{\frac{z_0}{\e}+{\rm const}+\order{\e}\Big\}
\;,
\la{eq:z0}
\quad z_0=\frac1{24}
\(\frac{\zeta(3)}5+\frac{\zeta'(-1)}{\zeta(-1)}-\gammaE-\frac32\)
\;,
\end{align}
where const is somewhat lengthy, so we do not display it here.

%
\subsection{Evaluation of $\cS_4$}
\la{se:4th}

Expanding $[\Pi_B-\Pi_D]=
G(1,1,d\!+\!1)[(P^2)^{-\e}\!-\!(\alpha_1 T^2)^{-\e}]
\approx G(1,1,4\!-\!2\e)\,\e\,[\ln\frac{\alpha_1 T^2}{P^2}+\order{\e}]$ 
as well as
$G(1,1,4-2\e)\approx\frac{(4\pi e^2/e^{\gammaEs})^\e}{(4\pi)^2\,\e}
\(1+\order{\e^2}\)$,
we see that this integral is finite,
so we write 
\begin{align}
\cS_4 &\equiv
\sumintp{P} \frac1{\lk P^2\rk^2} 
\lk\Pi_B-\Pi_D\rk\lk\bar\Pi-\bar\Pi_C\rk
\approx\frac{T^2}{(4\pi)^4}\,\sS_4+\order{\e}\;.
\end{align}
Using the 3d spatial Fourier transform of 
$[\bar\Pi-\bar\Pi_C]$ from \app\ref{se:FT};
integrating over angles via \eq\nr{eq:av3};
letting $|\vec{r}|=x/(2\pi T)$, $|\vec{p}|=|\pz|y$ and $|\pz|=2\pi Tn$;
and using the exponential-integral \eq\nr{eq:ei1}:
\begin{align}
\la{eq:S4}
\sS_4=&\,\frac16\sum_{n=1}^\infty \frac1n \intx 
\lb e^{-2nx}\ln\(\frac{2x}{n}\frac{\alpha_1 e^{\gammaEs-1}}{16\pi^2}\)
-{\rm Ei}(-2nx)\rb\Big[\bar f(x,n)-\bar f_C(x,n)\Big]\\
=&\,z_0\ln\(\frac{\alpha_1 e^{\gammaEs-1}}{16\pi^2}\)+\sS_{4a}\\
&\sS_{4a}=z_0\ln(\alpha_6)+\frac16\sum_{n=1}^\infty \frac1n \intx 
\lk e^{-2nx}\ln\(\frac{2x}{\alpha_6 n}\)
-{\rm Ei}(-2nx)\rk\Big[\bar f(x,n)-\bar f_C(x,n)\Big]\nn
&\hphantom{\sS_{4a}}\approx +0.004820184(1)\;,\la{eq:numV4a}
\end{align}
where the explicit $\alpha_1$\/-dependence follows from 
$\alpha_1$\/-independence of $\cS_3+\cS_4$ term with $z_0$ from \eq\nr{eq:z0},
and we have introduced an arbitrary parameter $\alpha_6$
to parameterize some freedom in evaluating the number $\sS_{4a}$.
Mathematica \cite{mma} actually gives a result for the sum for the first term 
in curly brackets of \eq\nr{eq:S4}, but none for the ${\rm Ei}$ term.
For obtaining the numerical approximation, we have truncated 
the sum over $n$ at some $n_{\rm max}=30000$.
In order to estimate the magnitude of the remainder, we have interpolated
the one-dimensional integral in the range $n \in [10^4,10^5]$ with a
simple power law $n^{-a}$ and performed the summation from
$n=30001\dots\infty$ analytically, from which we infer that the 
remainder is of $\order{10^{-10}}$.

%
\subsection{Evaluation of $\cS_5$}
\la{se:5th}

In close analogy to the treatment of $\cS_3$, 
we identically re-write the fifth term of \eq\nr{eq:V2rewritten} as
\begin{align}
\la{eq:s5b}
\cS_5&\equiv
\sumintp{P}\frac1{P^4}\lk\Pi\!-\!\Pi_C\rk\bar\Pi_D
=\frac{G(1,1,d\!+\!1)}{(\alpha_2 T^2)^{(3\!-\!d)/2}}\,\frac1{4d}
\sumint{P}\frac{(d\!+\!1)\Pz^2\!-\!P^2}{P^4}\lb
\Pi
\!-\!\dpzero\Pi
\!-\!\Pi_C
\!+\!\dpzero\Pi_C
\rb\;.
\end{align}
Now the first term of \eq\nr{eq:s5b} 
is a sum of two regular 2-loop sum-integrals, which 
both reduce to zero:
first, from the IBP relation of \eq\nr{eq:ibp2} we get
\begin{align*}
\sumint{P}\frac{(d+1)\Pz^2-P^2}{P^4}\Pi&
=(d+1)\,I(211,20)-I(111,00)
=-\frac{d(d-2)}3\,I(111,00)\;,
\end{align*}
while furthermore, the 2-loop sunset sum-integral $I(111,00)$ vanishes
identically via IBP identity \eq\nr{eq:ibp4}.
The second term of \eq\nr{eq:s5b} is a special
2-loop tadpole $A$; the third term is immediately recognized
to be a sum of 1-loop tadpoles by noticing that the $P$\/-dependence 
of $\Pi_C$ is trivial (powers of $P^2$ only);
and, for the same reason, 
the fourth term is scale-free and hence vanishes in dimensional 
regularization.
We hence get the exact expression
\begin{align}
\cS_5=&\,
\frac{G(1,1,d+1)}{(\alpha_2 T^2)^{(3-d)/2}}\,\frac1{4d}\Big\{
[0_{IBP}-0_{IBP}]-[0-A(1,1,1;0)]
-\nn&-
\lk G(1,1,d+1)\((d+1)I_{2+\e}^2-I_{1+\e}^0\)
+2I_1^0\((d+1)I_3^2-I_2^0\)\rk
+0_{\rm scalefree}
\Big\}
\\\approx&\,\frac{T^2}{(4\pi)^4}\(\frac1{\alpha_2 T^6}\)^\e \Big\{
\frac{z_1}{\e}
+{\rm const}+\order{\e}\Big\}
\;,\quad 
\la{eq:z1}
z_1=\frac1{12}\(\frac{\zeta'(-1)}{\zeta(-1)}-\ln(2\pi)-\frac16\)
\;,
\end{align}
where const is somewhat lengthy, so we do not display it here.

%
\subsection{Evaluation of $\cS_6$}
\la{se:6th}

In close analogy to the treatment of $\cS_4$, expanding 
$[\bar\Pi_B-\bar\Pi_D]
\approx\frac{(d+1)\Pz^2-P^2}{4d}\frac1{(4\pi)^2}
[\ln\frac{\alpha_2 T^2}{P^2}+\order{\e}]$ 
we see that the sixth term of \eq\nr{eq:V2rewritten} is finite, 
so we write 
\begin{align}
\cS_6 &\equiv
\sumintp{P} \frac1{\lk P^2\rk^2} 
\lk\Pi-\Pi_C\rk\lk\bar\Pi_B-\bar\Pi_D\rk
\approx\frac{T^2}{(4\pi)^4}\,\sS_6+\order{\e}\;.
\end{align}
Using the 3d spatial Fourier transform of 
$[\Pi-\Pi_C]$ from \app\ref{se:FT};
integrating over angles via \eq\nr{eq:av3};
letting $|\vec{r}|=x/(2\pi T)$, $|\vec{p}|=|\pz|y$ and $|\pz|=2\pi Tn$:
\begin{align}
\sS_6&=\frac1{3\pi}\sum_{n=1}^\infty \intx
\inty\,\frac{(3-y^2)}{(y^2+1)^2}
\ln\(\frac{\alpha_2/(2\pi n)^2}{y^2+1}\) 
y\sin(nxy)\frac{e^{-nx}}x\Big[f(x,n)-f_C(x,n)\Big]
\end{align}
Now, splitting $(3-y^2)=4-(y^2+1)$ and integrating over $y$ 
using \eqs\nr{eq:ei1}, \nr{eq:ei2}:
\begin{align}
\la{eq:s5a}
\sS_6=&\,
\frac13\sum_{n=1}^\infty \intx\,
n\lk e^{-2nx}(L-1)-{\rm Ei}(-2nx)\rk\Big[f(x,n)-f_C(x,n)\Big]
-\nn&-
\frac16\sum_{n=1}^\infty \intx\,
\frac1x\lk e^{-2nx}L+{\rm Ei}(-2nx)\rk\Big[f(x,n)-f_C(x,n)\Big]\;,
\end{align}
where $L\equiv\ln\(\frac{2x}{n}\frac{\alpha_2 e^\gammaEs}{16\pi^2}\)$.
At $\alpha_2=16\pi^2 e^{-\gammaEs}$, the last line of \eq\nr{eq:s5a} equals 
$\frac{C}{36}\approx\frac{0.003496}{36}$
(see \eq(2.14) of \cite{Schroder:2012hm}, 
where its numerical evaluation had cost some sweat;
see also \eq(D.27) of \cite{Andersen:2008bz}, where $0.0034814$ is quoted).
Noting that the $\alpha_2$\/-dependence of \eq\nr{eq:s5a} follows from 
$\alpha_2$\/-independence of $\cS_5+\cS_6$, let us re-write it as
(with $z_1$ from \eq\nr{eq:z1} above)
\begin{align}
\sS_6&=
z_1\ln\(\frac{\alpha_2 e^\gammaEs}{16\pi^2}\)+\sS_{6a}+\sS_{6b}+\frac{C}{36}
\approx z_1\ln\(\frac{\alpha_2 e^\gammaEs}{16\pi^2}\)+0.000037813(1)\;,\\
\sS_{6a}&=3\alpha_3 z_1+\intx\sum_{n=1}^\infty 
e^{-2nx}\Big[f(x,n)-f_C(x,n)\Big]
\lk\frac{\alpha_3}{2x}-n\(\alpha_3+\frac13\)\rk\nn
&=3\alpha_3 z_1+\frac14\intx\Big[f(x,0)-f_C(x,0)\Big]
\lk\alpha_3\frac{\coth(x)-1}x+\(\alpha_3+\frac13\)\(1-\coth^2(x)\)\rk\nn
&\approx +0.0025600539026700475010965(1)\;,\la{eq:numV6a} \\
\sS_{6b}&=-\sS_{6a}\ln(\alpha_4)-\frac13\intx\sum_{n=1}^\infty 
n\Big[f(x,n)-f_C(x,n)\Big]\lk{\rm Ei}(-2nx)-e^{-2nx}\ln\(\frac{2x}{\alpha_4 n}\)\rk\nn
&\approx -0.002619354(1)\;,\la{eq:numV6b}\\
C&=36(\sS_{6a}+z_1)\ln\(\alpha_5\)-6\intx\sum_{n=1}^\infty 
\frac1x\Big[f(x,n)-f_C(x,n)\Big]
\lk{\rm Ei}(-2nx)+e^{-2nx}\ln\(\frac{2x}{\alpha_5 n}\)\rk\nn
&\approx +0.0034960718(1)\la{eq:numC} \;,
\end{align}
and where we have introduced three arbitrary parameters 
$\{\alpha_3,\alpha_4,\alpha_5\}$
to parameterize some freedom we found in evaluating 
the three numbers $\{\sS_{6a},\sS_{6b},C\}$. 
Concerning numerical precision, see the discussion below \eq\nr{eq:numV4a}.
As mentioned above, the constant $C$ 
has already been evaluated  
in \cite{Schroder:2012hm} (at $\alpha_5=1$; see \app{B} therein).
Some further analytic work on these constants seems possible 
(for example, $f-f_C$ is $n$\/-independent and the corresponding sums 
can be performed; $\sS_{6a}$ seems to be a suitable 
candidate for analytic treatment), 
but the numeric result is at
this point fully sufficient for our purposes.

%
\subsection{Evaluation of $\cS_7$--$\cS_9$}
\la{se:7th8th9th}

The last three terms of \eq\nr{eq:V2rewritten} are trivial tadpoles, and 
can immediately be evaluated analytically 
in terms of the functions $G$ and $I$ from \app\ref{se:functions} as
\begin{align}
\cS_7&\equiv\sumintp{P} \frac1{\lk P^2\rk^2} 
\lk\Pi_C-\Pi_B\rk\bar\Pi_B
\;=\;2G(1,1,d+1) I_1^0\(\frac{d+1}{4d}I_{3+\e}^2-\frac1{4d}I_{2+\e}^0\)
\\&\approx\frac{T^2}{(4\pi)^4}\(\frac1{4\pi T^2}\)^{3\e}\Big\{
\frac1{48\e}
+{\rm const}+\order{\e}
\Big\} \;,\\
\cS_8&\equiv\sumintp{P} \frac1{\lk P^2\rk^2} 
\Pi_B\lk\bar\Pi_C-\bar\Pi_B\rk
\;=\;G(1,1,d+1)\(I_1^0 I_{3+\e}^2+2I_1^2 I_{3+\e}^0\)
\\&\approx\frac{T^2}{(4\pi)^4}\(\frac1{4\pi T^2}\)^{3\e}\Big\{
\frac1{96\e^2}
+\frac{\frac{13}2+\gammaE+2\frac{\zeta'(-1)}{\zeta(-1)}-\frac45\zeta(3)}{96\e}
+{\rm const}+\order{\e}
\Big\} \;,\\
\cS_9&\equiv\sumintp{P} \frac1{\lk P^2\rk^2} 
\Pi_B\bar\Pi_B
\;=\;G(1,1,d+1)^2\(\frac{d+1}{4d}I_{2+2\e}^2-\frac1{4d}I_{1+2\e}^0\)
\\&\approx\frac{T^2}{(4\pi)^4}\(\frac1{4\pi T^2}\)^{3\e}\Big\{
-\frac1{48\e^2}
+\frac{-5+3\gammaE-6\frac{\zeta'(-1)}{\zeta(-1)}}{48\e}
+{\rm const}+\order{\e}
\Big\} \;.
\end{align}

%
\section{Result}
\la{se:resultVb}

Collecting all expressions that have been evaluated in the previous
sections in order to re-assemble $\Vb$ according to \eq\nr{eq:V1to9}, 
setting $d=3-2\e$ and expanding, we obtain our final result
\begin{align}
\Vb &\approx \frac{T^2}{(4\pi)^4}\, 
\frac{\(4\pi T^2\)^{-3\e}}{96\,\e^2} 
\lk 1 +v_{21}\,\e +v_{22}\,\e^2 +\order{\e^3}\rk\;,\\
v_{21}&=\frac{67}{6}+\gammaE+2\frac{\zeta'(-1)}{\zeta(-1)}\;,\\
v_{22}&=
\frac{443}{36}
-\frac{39}2\gammaE^2
+\gammaE\(\frac{23}6+6\frac{\zeta'(-1)}{\zeta(-1)}-8\ln(2\pi)+\frac45\,\zeta(3)\)
-\frac65\,\zeta(3)
-\frac85\,\zeta'(3)
+\nn&+
\frac{143}{36}\,\pi^2
+8\ln^22+16\ln(2\pi)+8\ln(\pi)\ln(4\pi)
-48\gamma_1
+21\frac{\zeta'(-1)}{\zeta(-1)}
-6\frac{\zeta''(-1)}{\zeta(-1)}
+\nn&+
96\Big[-\frac13(\sS_{1a}+\sS_{1b})+\sS_{2a}+\sS_{4a}+\sS_{6a}
+\sS_{6b}+\frac{1}{36}\,C\Big]
\\
\la{eq:numRes}
&\approx 128.63807196263994701\dots+96\big[-0.370298231(2)\big]
=93.0894417(2) \;,
\end{align}
where the Stieltjes constant $\gamma_1$ is defined
by $\zeta(1+\e)=1/\e+\gammaE-\gamma_1\,\e+\order{\e^2}$,
and the constant $v_{22}$ contains the seven (sum-)integrals
for which we do not have analytic representations. They are defined
in \eqs\nr{eq:numV1a}, \nr{eq:numV1b}, \nr{eq:s1234}, 
\nr{eq:numV4a}, \nr{eq:numV6a}, \nr{eq:numV6b} and \nr{eq:numC},
and were evaluated numerically in the above.
The main uncertainty in the final numerical result \eq\nr{eq:numRes}
is dominated by the ones in $\sS_{4a}$ and $\sS_{6b}$.

%
\section{Conclusions}
\la{se:conclu}

We have successfully evaluated a new massless bosonic three-loop
sum-integral of mass dimension two, which we have named $\Vb$, 
and which represents one of the two 
remaining unknown ingredients for a 3-loop evaluation
of the Debye screening mass in hot Yang-Mills 
theory, \cite{Moeller:2012da,debyeMass}.

Our strategy has relied heavily on the particular structure of
$\Vb$, containing two different one-loop two-point sub-integrals,
whose known analytic properties we have repeatedly exploited.
This particular line of attack has served to evaluate a number
of similar (but somewhat less involved) cases of sum-integrals
in the past.

We have obtained all divergent terms of $\Vb$ analytically,
which will be important when it is finally used in a concrete 
physics computation; for the constant part, we had to resort
to a numerical treatment for some pieces thereof.
While it is certainly possible to increase the precision of our 
numerics substantially, we feel that the numbers given above will
be sufficiently accurate for all practical purposes. 
It would of course be nice to obtain all terms in
$v_{22}$ analytically, but we have no systematic method to do so 
for now.
In passing, there might be some relation of (parts of) $\Vb$ 
to the sum-integral $\intV$ that was treated
in \cite{Andersen:2008bz,Schroder:2012hm}, 
since the constant $C$ of our \eq\nr{eq:numC}
had already contributed there. However, his might also
be pure coincidence, and we have not pursued this relation further.

Finally, we hope that the last remaining sum-integral that is needed
at dimension two is amenable to
similar techniques, and leave its computation for future work.

\acknowledgments

The work of I.G.~has been supported by the Deutsche
Forschungsgemeinschaft (DFG) under grant no.~GRK~881.
Y.S.~is supported by the Heisenberg program of the DFG, 
contract no.~SCHR~993/1. 


\begin{appendix} 

%
\section{3d Fourier transforms}
\la{se:FT}

All (inverse) 3d spatial Fourier transforms derive from 
\eqs(22-24) in \cite{Moller:2010xw}:
\begin{align}
\frac1{(\vec{q}^2+\qz^2)^s}&=\frac{2\pi T^2}{(8\pi T^2)^s}
\int\frac{{\rm d}^3\vec{r}}{r}\,e^{i\vec{q}\vec{r}}\,e^{-|\qz|r}
\(\frac{\bar r}{|\bar \qz|}\)^{s-1} f_s(|\bar \qz|r)\;,\\
{\rm where}&\quad f_s(m)=\sqrt{\frac{2m}\pi}\,e^m\,K_{3/2-s}(m)\;,
\end{align}
with special cases of the modified Bessel function of second kind 
evaluating to $f_1(m)=f_2(m)=1$ and $f_0(m)=f_3(m)=(1+1/m)$. From 
this one gets a generic formula for e.g. 
(inverse) 3d spatial Fourier transforms of 1-loop $\Pi$'s:
\begin{align}
\Pi_{ij}^k(P)&\equiv\sumint{Q}\frac{\Qz^k}{[Q^2]^i[(P-Q)^2]^j}
=\frac{T(2\pi T)^k(2\pi T^2)^2}{(8\pi^2 T^2)^{i+j}}
\int\frac{{\rm d}^3\vec{r}}{r^2}\,e^{i\vec{p}\vec{r}}\,e^{-|\pz|r}\,
\pi_{ij}^k(\bar r,|\pzb|)\\
\pi_{ij}^k(\bar r,|\pzb|)&=
\!\sum_{n=-\infty}^\infty\!\! n^k\,e^{-(|n|+|\pzb\!-\!n|-|\pzb|)\bar r}
\(\frac{\bar r}{|n|}\)^{i-1}\! f_i(|n|\bar r)\,
\(\frac{\bar r}{|\pzb\!-\!n|}\)^{j-1} f_j(|\pzb\!-\!n|\bar r) \;.
\end{align}
Matsubara sums can then be solved using \eq(19) of \cite{Moller:2010xw},
as reprinted in \eq\nr{eq:sum19} below.
In particular, we get the (inverse) 3d spatial Fourier transforms
(see also \cite{Schroder:2012hm})
\begin{align}
\Pi_i(P)=&\frac{T}{(4\pi)^2}\int\frac{{\rm d}^3\vec r}{r^2}e^{i\vec p\vec
  r}e^{-|\pz|r} f_i(\bar r,|\bar \pz|)\\
&f(x,n)=\coth(x)+n\;,\\
&f_B(x,n)=n+\frac1x\;,\\
&f_C(x,n)=n+\frac1x+\frac{x}3\;,\\
\bar\Pi_i(P)=&\frac{T^3}{24}\int\frac{{\rm d}^3\vec r}{r^2}e^{i\vec p\vec
  r}e^{-|\pz|r} \bar{f}_i(\bar r,|\bar \pz|)\\
\la{eq:FTbf}
&\bar{f}(x,n)=3\coth^3(x)\!-\!3\coth(x)\!+\!n(3\coth^2(x)\!-\!2)
\!+\!3n^2\coth(x)\!+\!2n^3\;,
\\
\la{eq:FTbfB}
&\bar{f}_B(x,n)=\frac3{x^3}+\frac{3n}{x^2}+\frac{3n^2}{x}+2n^3\;,\\
&\bar{f}_C(x,n)=\frac3{x^3}+\frac{3n}{x^2}+\frac{3n^2}{x}+2n^3
+n^2 x-\frac{x}5\;,\\
\bar{\bar \Pi}_i(P)=&
\frac{T}{4(4\pi)^2}\int\frac{{\rm d}^3\vec r}{r^2}e^{i\vec p\vec r}
e^{-|\pz|r} \bar{\bar{f}}_i(\bar r,|\bar \pz|)\;,\\
&\bar{\bar{f}}(x,n)=x\lk\coth^2(x)-1+n\coth(x)+n^2\rk\;,\\
&\bar{\bar{f}}_B(x,n)=x\lk\frac1{x^2}+\frac{n}x+n^2\rk\;,\\
\dpzero\widetilde{\Pi}'(P)=&
\frac{4}{T(4\pi)^4}\,\dpzero \int\frac{{\rm d}^3\vec r}{r^2}e^{i\vec p\vec
  r}e^{-|\pz|r} \tilde f'(\bar r,0)\;,\\
\la{eq:FTtfp}
&\tilde f'(x,0)=-x\ln\(1-e^{-2x}\) \;.
\end{align}

%
\section{Standard integrals}
\la{se:functions}

For convenience, we collect here the functions used above, as defined
in \cite{Moller:2010xw,Schroder:2012hm}. 
They are the zero-temperature massless one-loop propagator 
\begin{align}
\la{eq:Gdef}
G(s_1,s_2,d) &\equiv \(p^2\)^{s_{12}-\frac{d}2}\int\frac{{\rm d}^dq}{(2\pi)^d}
\frac1{[q^2]^{s_1}[(q-p)^2]^{s_2}}
=
\frac{\Gamma(\frac{d}2-s_1)\Gamma(\frac{d}2-s_2)\Gamma(s_{12}-\frac{d}2)}
{(4\pi)^{d/2}\Gamma(s_1)\Gamma(s_2)\Gamma(d-s_{12})} \;;
\end{align}
the one-loop bosonic tadpoles 
\begin{align}
\la{eq:Idef}
I_s^a \equiv \sumint{Q} \frac{|\qz|^a}{[Q^2]^s} 
= \frac{2T\,\zeta(2s-a-d)}{(2\pi T)^{2s-a-d}}\,
\frac{\Gamma(s-\frac{d}2)}{(4\pi)^{d/2}\Gamma(s)}
\;,\quad
I_s \equiv \sumint{Q} \frac1{[Q^2]^s} 
= I_s^0\;;
\end{align}
and a specific two-loop tadpole
\begin{align}
\la{eq:Atad0}
A(s_1,s_2,s_3) &\equiv A(s_1,s_2,s_3;0) \;,\\
\la{eq:Atad}
A(s_1,s_2,s_3;s_4) &\equiv \sumint{PQ} 
\frac{\dqzero|\pz|^{s_4}}{[Q^2]^{s_1}[P^2]^{s_2}[(P-Q)^2]^{s_3}}
\nn&=
\frac{2T^2\,\zeta(2s_{123}-2d-s_4)}{(2\pi T)^{2s_{123}-2d-s_4}}\,
\frac{\Gamma(s_{13}-\frac{d}2)\Gamma(s_{12}-\frac{d}2)
\Gamma(\frac{d}2-s_1)\Gamma(s_{123}-d)}
{(4\pi)^d\Gamma(s_2)\Gamma(s_3)\Gamma(d/2)\Gamma(s_{1123}-d)}
\;,
\end{align}
where $s_{abc...} \equiv s_c+s_b+s_c+...\;$;

%
\section{Other integrals}
\la{se:Ei}

In the main text, we have used the exponential integral, defined as
\begin{align}
{\rm Ei}(x)\equiv -\int_{-x}^\infty \frac{{\rm d}t}t \,e^{-t}
\end{align}
for the following integrals (the first of which has already been
used in \se 2.2 of \cite{Schroder:2012hm}):
\begin{align}
\la{eq:ei2}
\frac2\pi\,e^{|z|}\inty\,\frac{y\,\sin(y|z|)}{y^2+1}\,
\ln\frac{\alpha}{y^2+1}&=
e^{2|z|}{\rm Ei}(-2|z|)+\gammaE+\ln\frac{|z|\alpha}{2}\;,\\
\la{eq:ei1}
\frac{4\,e^{|z|}}{\pi |z|}\inty\,
\frac{y\,\sin(y|z|)}{(y^2+1)^2}\,
\ln\frac{\alpha}{y^2+1}&=
\ln\frac{|z|\alpha}{2e^{1-\gammaEs}}-e^{2|z|}{\rm Ei}(-2|z|)\;.
\end{align}

Furthermore, we needed one special 3d Fourier transform 
and its generalization
\begin{align}
\la{eq:av1}
\int\frac{{\rm d}^3\vec{p}}{(2\pi)^3}\,
e^{i\vec p \vec r}\frac1{(\vec p^2+\pz^2)^2}&=
\frac{e^{-|\pz|r}}{8\pi|\pz|}\;,\\
\la{eq:av4}
\int\frac{{\rm d}^3\vec{p}}{(2\pi)^3}\,
e^{i\vec p \vec r}\frac1{(\vec p^2+\pz^2)^{n+1}}&=
\frac1{n!}\partial_{-\pz^2}^{\,n}\frac{e^{-|\pz|r}}{4\pi r}=
\frac{e^{-|\pz|r}}{4\pi}\frac{2|\pz|}{(2|\pz|)^{2n}}f_n(|\pz|r)\;,\\
\mbox{with}&\quad f_{\{0,1,2,3\}}(x)=\{1/(2x),1,1+x,2+2x+2x^2/3\}\;,
\end{align}
as well as the following angular averages:
\begin{align}
\la{eq:av3}
\frac12\int_{-1}^1\!\!\!\!{\rm d}u\,e^{i p r u}&=
\frac{\sin(pr)}{pr}\;,\\
\la{eq:av2}
\frac12\int_{-1}^1\!\!\!\!{\rm d}u\,e^{-n\sqrt{x^2+y^2+2xyu}}&=
\frac1{2xyn^2}\lk(1\!+\!n|x\!-\!y|)e^{-n|x-y|}-(1\!+\!n(x\!+\!y))e^{-n(x+y)}\rk\;,\\
\la{eq:av5}
\frac12\int_{-1}^1\!\!\!\!{\rm d}u\,\frac1{\sqrt{x^2+y^2+2xyu}}&=
\frac{x+y-|x-y|}{2xy}\;.
\end{align}

%
\section{Summation formulae}
\la{se:sums}

When dealing with discrete sums (see also \cite{Moller:2010xw}),
Zeta functions and polylogarithms enter through 
\begin{align}
\zeta(s)&\equiv \sum_{n=1}^\infty n^{-s}\;,
\la{eq:sum1}
&\mbox{Li}_s(x) \equiv \sum_{n=1}^\infty \frac{x^n}{n^s} \,
\end{align}
with
$\mbox{Li}_1(x)=-\ln(1-x)$
(one can use \eq\nr{eq:sum1} at $\{s=1,x=e^{-r}\}$ and then 
get the cases $s<1$ via $\partial_r$),
while hyperbolic functions appear via ($m\in\mathbb{Z}$)
\begin{align}
\la{eq:sum18}
\sum_{n=-\infty}^\infty e^{-(|n|+|n-m|-|m|)r}
&= |m|+\coth(r) \;,\\
\la{eq:sum19}
\sum_{n=-\infty}^\infty e^{-(|n|+|n-m|-|m|)r} p(|n|)
&=\sum_{n=0}^{|m|}p(n)
+p(-\partial_{2r})\frac1{e^{2r}\!-\!1}
+e^{2|m|r}p(-\partial_{2r})\frac{e^{-2|m|r}}{e^{2r}\!-\!1}\;,
\end{align}
where $p(x)$ can be a polynomial. 
Note that \eq\nr{eq:sum18} is
just a special case of \eq\nr{eq:sum19}.

%
\section{IBP relations}
\la{se:ibp}

We need one 3-loop IBP relation for \se\ref{se:1st},
which can be derived as
\begin{align}
\la{eq:ibp1}
0=&\,\partial_{\vec p} \vec p\circ \cS_1\nn
=&\,d\,\cS_1+\sumint{PQR}\vec p\partial_{\vec p}
\frac{\dpzero\,\Rz^2}{[P^2]^2 Q^2(P-Q)^2 R^2(P-R)^2}\nn
=&\,d\,\cS_1+\sumint{PQR}
\frac{\dpzero\,\Rz^2 \cdot N}{[P^2]^3 Q^2[(P-Q)^2]^2 R^2[(P-R)^2]^2}\nn
&\quad \dpzero N=\dpzero\big\{
-4{\vec{p}\vec{p}}(P\!-\!Q)^2(P-R)^2
-{2\vec{p}(\vec{p}\!-\!\vec{q})}P^2(P-R)^2
-{2\vec{p}(\vec{p}\!-\!\vec{r})}P^2(P-Q)^2\big\}\nn
&\qquad\qquad\;\;\mbox{let~~}
\vec{p}\vec{p}\rightarrow{P^2},\;
{2\vec{p}(\vec{p}\!-\!\vec{q})}\rightarrow{(P\!-\!Q)^2\!-\!Q^2\!+\!P^2},\;
{2\vec{p}(\vec{p}\!-\!\vec{r})}\rightarrow{(P\!-\!R)^2\!-\!R^2\!+\!P^2}
\nn
&\quad \hphantom{\dpzero N}=\dpzero P^2\big\{ 
-6(P-Q)^2(P-R)^2
+(Q^2-P^2)(P-R)^2
+(R^2-P^2)(P-Q)^2\big\}\nn
&\quad\mbox{shift $Q\rightarrow P-Q$ and $R\rightarrow P-R$ }\nn
=&\,(d-6)\cS_1+\sumint{P}\frac{\dpzero}{[P^2]^2}
\(\sumint{Q}\frac{(P-Q)^2-P^2}{[Q^2]^2(P-Q)^2}\,\bar\Pi
+\Pi\sumint{R}\frac{\Rz^2((P-R)^2-P^2)}{[R^2]^2(P-R)^2}\) \;.
\end{align}

The 2-loop IBP relations that we need 
for \se\ref{se:3rd} and \se\ref{se:5th}
can be obtained in the same way.
Letting
$I(abc,de)\equiv\sint_{PQ}\frac{[\Pz]^d[\Qz]^e}{[P^2]^a[Q^2]^b[(P-Q)^2]^c}$,
we can derive e.g.\
\begin{align}
\la{eq:ibp2}
0&=\partial_{\vec{p}}\vec{p}\circ I(111,00)
\nn&
=d\,I(111,00)+\sumint{PQ}\partial_{\vec{p}}\vec{p}\frac1{P^2Q^2(P-Q)^2}
\nn&
=d\,I(111,00)+\sumint{PQ}\(\frac{-2\vec{p}\vec{p}}{[P^2]^2Q^2(P-Q)^2}
+\frac{-2\vec{p}(\vec{p}-\vec{q})}{P^2Q^2[(P-Q)^2]^2}\)
\;\mbox{do $P\rightarrow Q-P$ in last term}
\nn&
=d\,I(111,00)+\sumint{PQ}\frac{-4\vec{p}^2+2\vec{p}\vec{q}}{[P^2]^2Q^2(P-Q)^2}
\nn&
=d\,I(111,00)+\sumint{PQ}\frac{-3(P^2-\Pz^2)-\vec{p}(\vec{p}-2\vec{q})}
{[P^2]^2Q^2(P-Q)^2}
\;\mbox{~~last term odd under $Q\rightarrow P-Q$}
\nn&
=(d-3)\,I(111,00)+3\,I(211,20) \;.
\end{align}
More conveniently, using our computer-algebraic \cite{Vermaseren:2000nd}
implementation of a Laporta-type \cite{Laporta:2001dd} algorithm
to systematically derive and solve such 
finite-temperature IBP relations \cite{Nishimura:2012ee},
we get the special cases
\begin{align}
\la{eq:ibp4}
I(111;00)&=0\;,\quad\mbox{(for explicit IBP derivation see \cite{Schroder:2008ex})}\\
\la{eq:ibp3}
I(211;20)&=0\;,\\
\la{eq:ibp5}
I(211;02)&=-\frac1{d-5}\(2I_1^0 I_3^2+I_2^2 I_2^0\)\;,
\end{align}
where the second relation confirms \eq\nr{eq:ibp2}.


\end{appendix}




\begin{thebibliography}{99}


\bibitem{Moeller:2012da}
  J.~M\"oller and Y.~Schr\"oder,
  {\em Three-loop matching coefficients for hot QCD: Reduction and gauge independence},
  arXiv:1207.1309.

\bibitem{Laporta:2001dd}
  S.~Laporta,
  {\em High precision calculation of multiloop Feynman integrals by difference equations},
  Int.\ J.\ Mod.\ Phys.\ A {\bf 15} (2000) 5087
  [hep-ph/0102033].

\bibitem{Chetyrkin:1981qh}
  K.~G.~Chetyrkin and F.~V.~Tkachov,
  {\em Integration by Parts: The Algorithm to Calculate beta Functions in 4 Loops},
  Nucl.\ Phys.\ B {\bf 192} (1981) 159;
  F.~V.~Tkachov,
  {\em A Theorem on Analytical Calculability of Four Loop Renormalization Group Functions},
  Phys.\ Lett.\ B {\bf 100} (1981) 65.

\bibitem{Nishimura:2012ee}
  M.~Nishimura and Y.~Schr\"oder,
  {\em IBP methods at finite temperature},
  arXiv:1207.4042.

\bibitem{Gynther:2007bw}
  A.~Gynther, M.~Laine, Y.~Schr\"oder, C.~Torrero and A.~Vuorinen,
  {\em Four-loop pressure of massless O(N) scalar field theory},
  JHEP {\bf 0704} (2007) 094
  [hep-ph/0703307].

\bibitem{Schroder:2008ex}
  Y.~Schr\"oder,
  {\em Loops for Hot QCD},
  Nucl.\ Phys.\ Proc.\ Suppl.\  {\bf 183B} (2008) 296
  [arXiv:0807.0500].

\bibitem{Moller:2010xw}
  J.~M\"oller and Y.~Schr\"oder,
  {\em Open problems in hot QCD},
  Nucl.\ Phys.\ Proc.\ Suppl.\  {\bf 205-206} (2010) 218
  [arXiv:1007.1223].

\bibitem{debyeMass}
 I.~Ghi\c{s}oiu, J.~M\"oller and Y.~Schr\"oder,
 in preparation.

\bibitem{Andersen:2008bz}
  J.~O.~Andersen and L.~Kyllingstad,
  {\em Four-loop Screened Perturbation Theory},
  Phys.\ Rev.\ D {\bf 78} (2008) 076008
  [arXiv:0805.4478].

\bibitem{Schroder:2012hm}
  Y.~Schr\"oder,
  {\em A fresh look on three-loop sum-integrals},
  arXiv:1207.5666.

\bibitem{Arnold:1994ps}
  P.~B.~Arnold and C.~X.~Zhai,
  {\em The Three loop free energy for pure gauge QCD},
  Phys.\ Rev.\ D {\bf 50} (1994) 7603
  [hep-ph/9408276].

\bibitem{mma}
 Wolfram Research, Inc., {\tt Mathematica}, Version 8.0, 
 Champaign, IL (2012).

\bibitem{Vermaseren:2000nd}
  J.~A.~M.~Vermaseren,
  {\em New features of FORM},
  math-ph/0010025;
  J.~Kuipers, T.~Ueda, J.~A.~M.~Vermaseren and J.~Vollinga,
  {\em FORM version 4.0},
  arXiv:1203.6543.


\end{thebibliography}
\end{document}